\documentstyle[12pt]{article}
\author{Lawrence M. Krauss\thanks{also Department of Astronomy. Research
supported in part by
the NSF, DOE and TRNLC.  Bitnet: Krauss@Yalehep}  and Soo-Jong
Rey**\thanks{Yale-Brookhaven
SSC Fellow. Bitnet: Soo@Yalehep}\\
\it{Institute for Theoretical Physics}\\
\it {University of California, Santa Barbara CA 93106**}\\
\it{\&}\\
\it{Center for Theoretical Physics, Sloane Laboratory}\\
 \it{Yale University, New Haven CT 06511 USA}}
\title{{Spontaneous CP Violation}\\
{at the Electroweak Scale}}
\date{February 1992}

\begin{document}
%
\def\journal{\topmargin .3in    \oddsidemargin .5in
        \headheight 0pt \headsep 0pt
        \textwidth 5.625in 
        \textheight 8.25in 
        \marginparwidth 1.5in
        \parindent 2em
        \parskip .5ex plus .1ex         \jot = 1.5ex}
%
\journal
\def\baselinestretch{1.2}
\catcode`\@=11
\def\marginnote#1{}
%
\newcount\hour
\newcount\minute
\newtoks\amorpm
\hour=\time\divide\hour by60
\minute=\time{\multiply\hour by60 \global\advance\minute by-\hour}
\edef\standardtime{{\ifnum\hour<12 \global\amorpm={am}%
        \else\global\amorpm={pm}\advance\hour by-12 \fi
        \ifnum\hour=0 \hour=12 \fi
        \number\hour:\ifnum\minute<10 0\fi\number\minute\the\amorpm}}
\edef\militarytime{\number\hour:\ifnum\minute<10 0\fi\number\minute}
\def\draftlabel#1{{\@bsphack\if@filesw {\let\thepage\relax
   \xdef\@gtempa{\write\@auxout{\string
      \newlabel{#1}{{\@currentlabel}{\thepage}}}}}\@gtempa
   \if@nobreak \ifvmode\nobreak\fi\fi\fi\@esphack}
        \gdef\@eqnlabel{#1}}
\def\@eqnlabel{}
\def\@vacuum{}
\def\draftmarginnote#1{\marginpar{\raggedright\scriptsize\tt#1}}
\def\draft{\oddsidemargin -.5truein
        \def\@oddfoot{\sl preliminary draft \hfil
        \rm\thepage\hfil\sl\today\quad\militarytime}
        \let\@evenfoot\@oddfoot \overfullrule 3pt
        \let\label=\draftlabel
        \let\marginnote=\draftmarginnote
   \def\@eqnnum{(\theequation)\rlap{\kern\marginparsep\tt\@eqnlabel}%
\global\let\@eqnlabel\@vacuum}  }
%
%
\def\preprint{\twocolumn\sloppy\flushbottom\parindent 2em
        \leftmargini 2em\leftmarginv .5em\leftmarginvi .5em
        \oddsidemargin -.5in    \evensidemargin -.5in
        \columnsep .4in \footheight 0pt
        \textwidth 10in \topmargin  -.4in
        \headheight 12pt \topskip .4in
        \textheight 7.1in \footskip 0pt
        \def\@oddhead{\thepage\hfil\addtocounter{page}{1}\thepage}
        \let\@evenhead\@oddhead \def\@oddfoot{} \def\@evenfoot{} }
%
\def\proceedings{\pagestyle{empty}
        \oddsidemargin .26in \evensidemargin .26in
        \topmargin .27in        \textwidth 145mm
        \parindent 12mm \textheight 225mm
        \headheight 0pt \headsep 0pt
        \footskip 0pt   \footheight 0pt}
%
\def\numberbysection{\@addtoreset{equation}{section}
        \def\theequation{\thesection.\arabic{equation}}}
\def\underline#1{\relax\ifmmode\@@underline#1\else
        $\@@underline{\hbox{#1}}$\relax\fi}
\def\titlepage{\@restonecolfalse\if@twocolumn\@restonecoltrue\onecolumn
     \else \newpage \fi \thispagestyle{empty}\c@page\z@
        \def\thefootnote{\fnsymbol{footnote}} }
\def\endtitlepage{\if@restonecol\twocolumn \else \newpage \fi
        \def\thefootnote{\arabic{footnote}}
        \setcounter{footnote}{0}}  
\catcode`@=12
\relax
%
%
\def\figcap{\section*{Figure Captions\markboth
        {FIGURECAPTIONS}{FIGURECAPTIONS}}\list
        {Figure \arabic{enumi}:\hfill}{\settowidth\labelwidth{Figure
999:}
        \leftmargin\labelwidth
        \advance\leftmargin\labelsep\usecounter{enumi}}}
\let\endfigcap\endlist \relax
\def\tablecap{\section*{Table Captions\markboth
        {TABLECAPTIONS}{TABLECAPTIONS}}\list
        {Table \arabic{enumi}:\hfill}{\settowidth\labelwidth{Table 999:}
        \leftmargin\labelwidth
        \advance\leftmargin\labelsep\usecounter{enumi}}}
\let\endtablecap\endlist \relax
\def\reflist{\section*{References\markboth
        {REFLIST}{REFLIST}}\list
        {[\arabic{enumi}]\hfill}{\settowidth\labelwidth{[999]}
        \leftmargin\labelwidth
        \advance\leftmargin\labelsep\usecounter{enumi}}}
\let\endreflist\endlist \relax
%
%
%
%
\makeatletter
\newcounter{pubctr}
\def\publist{\@ifnextchar[{\@publist}{\@@publist}}
\def\@publist[#1]{\list
        {[\arabic{pubctr}]\hfill}{\settowidth\labelwidth{[999]}
        \leftmargin\labelwidth
        \advance\leftmargin\labelsep
        \@nmbrlisttrue\def\@listctr{pubctr}
        \setcounter{pubctr}{#1}\addtocounter{pubctr}{-1}}}
\def\@@publist{\list
        {[\arabic{pubctr}]\hfill}{\settowidth\labelwidth{[999]}
        \leftmargin\labelwidth
        \advance\leftmargin\labelsep
        \@nmbrlisttrue\def\@listctr{pubctr}}}
\let\endpublist\endlist \relax
\makeatother
%
\def\mdot{\hskip -.1cm \cdot \hskip -.1cm}
%
\def\slash{\not\!}
%
%
\catcode`\@=11
\def\section{\@startsection {section}{1}{0pt}{-3.5ex plus -1ex minus
 -.2ex}{2.3ex plus .2ex}{\raggedright\large\bf}}
\catcode`\@=12
%
\def\mbf#1{\hbox{\boldmath $#1$}}
%
\newskip\humongous \humongous=0pt plus 1000pt minus 1000pt
\def\caja{\mathsurround=0pt}
\def\eqalign#1{\,\vcenter{\openup1\jot \caja
        \ialign{\strut \hfil$\displaystyle{##}$&$
        \displaystyle{{}##}$\hfil\crcr#1\crcr}}\,}
\newif\ifdtup
\def\panorama{\global\dtuptrue \openup1\jot \caja
        \everycr{\noalign{\ifdtup \global\dtupfalse
        \vskip-\lineskiplimit \vskip\normallineskiplimit
        \else \penalty\interdisplaylinepenalty \fi}}}
\def\eqalignno#1{\panorama \tabskip=\humongous
        \halign to\displaywidth{\hfil$\displaystyle{##}$
        \tabskip=0pt&$\displaystyle{{}##}$\hfil
        \tabskip=\humongous&\llap{$##$}\tabskip=0pt
        \crcr#1\crcr}}
\def\oldrefledge{\hangindent3\parindent}
\def\oldreffmt#1{\rlap{[#1]} \hbox to 2\parindent{}}
\def\oldref#1{\par\noindent\oldrefledge \oldreffmt{#1}
        \ignorespaces}
\def\figledge{\hangindent=1.25in}
\def\figfmt#1{\rlap{Figure {#1}} \hbox to 1in{}}
\def\fig#1{\par\noindent\figledge \figfmt{#1}
        \ignorespaces}
%
%
\def\ie{\hbox{\it i.e.}}        \def\etc{\hbox{\it etc.}}
\def\eg{\hbox{\it e.g.}}        \def\cf{\hbox{\it cf.}}
\def\etal{\hbox{\it et al.}}
\def\dash{\hbox{---}}
\def\cok{\mathop{\rm cok}}
\def\tr{\mathop{\rm tr}}
\def\Tr{\mathop{\rm Tr}}
\def\Im{\mathop{\rm Im}}
\def\Re{\mathop{\rm Re}}
\def\bR{\mathop{\bf R}}
\def\bC{\mathop{\bf C}}
\def\lie{\hbox{\it \$}} 
\def\partder#1#2{{\partial #1\over\partial #2}}
\def\secder#1#2#3{{\partial^2 #1\over\partial #2 \partial #3}}
\def\bra#1{\left\langle #1\right|}
\def\ket#1{\left| #1\right\rangle}
\def\VEV#1{\left\langle #1\right\rangle}
\let\vev\VEV
\def\gdot#1{\rlap{$#1$}/}
\def\abs#1{\left| #1\right|}
\def\pri#1{#1^\prime}
\def\ltap{\raisebox{-.4ex}{\rlap{$\sim$}} \raisebox{.4ex}{$<$}}
\def\gtap{\raisebox{-.4ex}{\rlap{$\sim$}} \raisebox{.4ex}{$>$}}
\def\contract{\makebox[1.2em][c]{
        \mbox{\rule{.6em}{.01truein}\rule{.01truein}{.6em}}}}
\def\half{{1\over 2}}
\def\beq{\begin{equation}}
\def\eeq{\end{equation}}
\def\ul{\underline}
\def\bea{\begin{eqnarray}}
\def\lrover#1{
        \raisebox{1.3ex}{\rlap{$\leftrightarrow$}} \raisebox{
0ex}{$#1$}}
%
\def\com#1#2{
        \left[#1, #2\right]}
\def\eea{\end{eqnarray}}
%
\def\bentarrow{\:\raisebox{1.3ex}{\rlap{$\vert$}}\!\rightarrow}
\def\longbent{\:\raisebox{3.5ex}{\rlap{$\vert$}}\raisebox{1.3ex}%
        {\rlap{$\vert$}}\!\rightarrow}
\def\onedk#1#2{
        \begin{equation}
        \begin{array}{l}
         #1 \\
         \bentarrow #2
        \end{array}
        \end{equation}
                }
\def\dk#1#2#3{
        \begin{equation}
        \begin{array}{r c l}
        #1 & \rightarrow & #2 \\
         & & \bentarrow #3
        \end{array}
        \end{equation}
                }
\def\dkp#1#2#3#4{
        \begin{equation}
        \begin{array}{r c l}
        #1 & \rightarrow & #2#3 \\
         & & \phantom{\; #2}\bentarrow #4
        \end{array}
        \end{equation}
                }
\def\bothdk#1#2#3#4#5{
        \begin{equation}
        \begin{array}{r c l}
        #1 & \rightarrow & #2#3 \\
         & & \:\raisebox{1.3ex}{\rlap{$\vert$}}\raisebox{-
0.5ex}{$\vert$}%
        \phantom{#2}\!\bentarrow #4 \\
         & & \bentarrow #5
        \end{array}
        \end{equation}
                }
%
\hyphenation{anom-a-ly}
\hyphenation{comp-act-ifica-tion}
%
%
\def\ap#1#2#3{           {\it Ann. Phys. (NY) }{\bf #1}, #2 (19#3)}
\def\apj#1#2#3{          {\it Astrophys. J. }{\bf #1}, #2 (19#3)}
\def\apjl#1#2#3{         {\it Astrophys. J. Lett. }{\bf #1}, #2 (19#3)}
\def\app#1#2#3{          {\it Acta Phys. Polon. }{\bf #1}, #2 (19#3)}
\def\ar#1#2#3{     {\it Ann. Rev. Nucl. and Part. Sci. }{\bf #1}, #2
(19#3)}
\def\com#1#2#3{          {\it Comm. Math. Phys. }{\bf #1}, #2 (19#3)}
\def\ib#1#2#3{           {\it ibid. }{\bf #1}, #2 (19#3)}
\def\nat#1#2#3{          {\it Nature (London) }{\bf #1}, #2 (19#3)}
\def\nc#1#2#3{           {\it Nuovo Cim.  }{\bf #1}, #2 (19#3)}
\def\np#1#2#3{           {\it Nucl. Phys. }{\bf #1}, #2 (19#3)}
\def\pl#1#2#3{           {\it Phys. Lett. }{\bf #1}, #2 (19#3)}
\def\pr#1#2#3{           {\it Phys. Rev. }{\bf #1}, #2 (19#3)}
\def\prep#1#2#3{         {\it Phys. Rep. }{\bf #1}, #2 (19#3)}
\def\prl#1#2#3{          {\it Phys. Rev. Lett. }{\bf #1}, #2 (19#3)}
\def\pro#1#2#3{          {\it Prog. Theor. Phys. }{\bf #1}, #2 (19#3)}
\def\rmp#1#2#3{          {\it Rev. Mod. Phys. }{\bf #1}, #2 (19#3)}
\def\sp#1#2#3{           {\it Sov. Phys.-Usp. }{\bf #1}, #2 (19#3)}
\def\sjn#1#2#3{           {\it Sov. J. Nucl. Phys. }{#1}, #2 (19#3)}
\def\srv#1#2#3{           {\it Surv. High Energy Phys. }{#1}, #2 (19#3)}
\def\tp{these proceedings}
\def\zp#1#2#3{           {\it Zeit. fur Physik }{\bf #1}, #2 (19#3)}
%
%
%
%
\catcode`\@=11
\def\eqnarray{\stepcounter{equation}\let\@currentlabel=\theequation
\global\@eqnswtrue
\global\@eqcnt\z@\tabskip\@centering\let\\=\@eqncr
\gdef\@@fix{}\def\eqno##1{\gdef\@@fix{##1}}%
$$\halign to \displaywidth\bgroup\@eqnsel\hskip\@centering
  $\displaystyle\tabskip\z@{##}$&\global\@eqcnt\@ne
  \hskip 2\arraycolsep \hfil${##}$\hfil
  &\global\@eqcnt\tw@ \hskip 2\arraycolsep
$\displaystyle\tabskip\z@{##}$\hfil
   \tabskip\@centering&\llap{##}\tabskip\z@\cr}

\def\@@eqncr{\let\@tempa\relax
    \ifcase\@eqcnt \def\@tempa{& & &}\or \def\@tempa{& &}
      \else \def\@tempa{&}\fi
     \@tempa
\if@eqnsw\@eqnnum\stepcounter{equation}\else\@@fix\gdef\@@fix{}\fi
     \global\@eqnswtrue\global\@eqcnt\z@\cr}

\catcode`\@=12
%
%
\font\tenbifull=cmmib10 
\font\tenbimed=cmmib10 scaled 800
\font\tenbismall=cmmib10 scaled 666
\textfont9=\tenbifull \scriptfont9=\tenbimed
\scriptscriptfont9=\tenbismall
\def\bmit{\fam9 }
\def\boldalpha{\fam=9{\mathchar"710B } }
\def\boldbeta{\fam=9{\mathchar"710C } }
\def\boldgamma{\fam=9{\mathchar"710D } }
\def\bolddelta{\fam=9{\mathchar"710E } }
\def\boldepsilon{\fam=9{\mathchar"710F } }
\def\boldzeta{\fam=9{\mathchar"7110 } }
\def\boldeta{\fam=9{\mathchar"7111 } }
\def\boldtheta{\fam=9{\mathchar"7112 } }
\def\boldiota{\fam=9{\mathchar"7113 } }
\def\boldkappa{\fam=9{\mathchar"7114 } }
\def\boldlambda{\fam=9{\mathchar"7115 } }
\def\boldmu{\fam=9{\mathchar"7116 } }
\def\boldnu{\fam=9{\mathchar"7117 } }
\def\boldomicron{\fam=9{\mathchar"716F } }
\def\boldxi{\fam=9{\mathchar"7118 } }
\def\boldpi{\fam=9{\mathchar"7119 } }
\def\boldrho{\fam=9{\mathchar"711A } }
\def\boldsigma{\fam=9{\mathchar"711B } }
\def\boldtau{\fam=9{\mathchar"711C } }
\def\boldupsilon{\fam=9{\mathchar"711D } }
\def\boldphi{\fam=9{\mathchar"711E } }
\def\boldchi{\fam=9{\mathchar"711F } }
\def\boldpsi{\fam=9{\mathchar"7120 } }
\def\boldomega{\fam=9{\mathchar"7121 } }
\def\boldvarepsilon{\fam=9{\mathchar"7122 } }
\def\boldvartheta{\fam=9{\mathchar"7123 } }
\def\boldvarpi{\fam=9{\mathchar"7124 } }
\def\boldvarrho{\fam=9{\mathchar"7125 } }
\def\boldvarsigma{\fam=9{\mathchar"7126 } }
\def\boldvarphi{\fam=9{\mathchar"7127 } }
\def\boldGamma{\fam=6{\mathchar"7000 } }
\def\boldDelta{\fam=6{\mathchar"7001 } }
\def\boldTheta{\fam=6{\mathchar"7002 } }
\def\boldLambda{\fam=6{\mathchar"7003 } }
\def\boldXi{\fam=6{\mathchar"7004 } }
\def\boldPi{\fam=6{\mathchar"7005 } }
\def\boldSigma{\fam=6{\mathchar"7006 } }
\def\boldUpsilon{\fam=6{\mathchar"7007 } }
\def\boldPhi{\fam=6{\mathchar"7008 } }
\def\boldPsi{\fam=6{\mathchar"7009 } }
\def\boldOmega{\fam=6{\mathchar"700A } }
\def\boldmitOmega{\fam=9{\mathchar"700A } }
\def\boldmitGamma{\fam=9{\mathchar"7000 } }
\def\boldmitDelta{\fam=9{\mathchar"7001 } }
\def\boldmitTheta{\fam=9{\mathchar"7002 } }
\def\boldmitLambda{\fam=9{\mathchar"7003 } }
\def\boldmitXi{\fam=9{\mathchar"7004 } }
\def\boldmitPi{\fam=9{\mathchar"7005 } }
\def\boldmitSigma{\fam=9{\mathchar"7006 } }
\def\boldmitUpsilon{\fam=9{\mathchar"7007 } }
\def\boldmitPhi{\fam=9{\mathchar"7008 } }
\def\boldmitPsi{\fam=9{\mathchar"7009 } }
\def\boldmitOmega{\fam=9{\mathchar"700A } }

\relax

\def\double{
        \renewcommand{\baselinestretch}{2}
        \large
        \normalsize
        }

\def\single {
                \renewcommand{\baselinestretch}{1}
                \large
                \normalsize
                }


\maketitle
\begin{picture}(0,0)(0,0)
\put(300,320){NSF-ITP-92-03}
\put(310,300){YCTP-P9-92}
\end{picture}
\vspace{-5pt}

\begin{abstract}

Utilizing results on the cosmology of anomalous discrete
symmetries we show that models of spontaneous CP violation
can in principle avoid the domain wall problem first pointed
out by Zel'dovich, Kobzarev and Okun.  A small but
nonzero $\theta_{QCD}$ explicitly breaks CP and can lift the
degeneracy of the two CP conjugate vacua through
nonperturbative effects so that the domain walls become
unstable, but survive to cosmologically interesting epochs.
We explore the viability of spontaneous CP violation in the
context of two Higgs models, and find that the
invisible axion solution of the strong CP problem cannot be
implemented without further extensions of the Higgs sector.
 \end{abstract}

\endtitlepage

Since the ground-breaking experiments on $K^0$-decay
in the 1960's, it has been recognized
that the weak interaction violates CP invariance
(and thus, assuming CPT, T as well) \cite{lincolnreview}.
Nevertheless, in the intervening three decades the mechanism
responsible for (flavor non-diagonal) CP violation has not
yet been conclusively elucidated.
Moreover, the recognition that weak CP violation
is communicated to the strong
interaction via the
QCD axial anomaly has confused the issue
further--especially with
the lack of any observable electric dipole moments
for the neutron and electron.

A very simple possibility is that CP invariance is
\sl spontaneously broken \rm in
conjunction with the breaking of other continuous
global and/or gauge symmetries.
T.D. Lee \cite{tdlee}
was the first to point out that this mechanism
is indeed possible through
a complex vacuum expectation value (VEV) of
Higgs fields in a two Higgs doublet model of the
$SU_L (2) \times U_Y (1)$ electroweak
interaction. His model gives rise to a flavor-changing
neutral Higgs boson exchange accompanied by the
spontaneously broken CP invariances. This leads to,
for example, $\Delta S = 2$ interactions at the tree level.

In order to suppress flavor-changing neutral Higgs exchange
interactions \cite{glashowweinberg},
Weinberg \cite{weinberg} proposed a class of
multi-Higgs models. In this case, CP invariance
may be broken either spontaneously through complex
Higgs VEVs or explicitly through complex-valued Higgs
self-coupling constants (or both).
Complex valued VEVs can also
result naturally in theories without
fundamental Higgs particles.
For example, in technicolor models, a complex-valued
vacuum misalignment \cite{preskill} of techniquark
bilinear condensates \cite{dashen} can occur
at the electroweak scale.

Of course, the simplest model of explicit CP breaking
through the Higgs couplings resulting in the
Cabbibo-Kobayashi-Maskawa (CKM) mass matrix is in good
agreement with current CP violation phenomonology.
Nevertheless the idea that one might be forced
beyond this minimal model has been revived with the
recognition that the baryon number of the
universe might be generated non-perturbatively at
temperatures characteristic of the weak
symmetry breaking scale, i.e. the electroweak
baryogenesis scenario \cite{ewbaryogenesis}. In this case,
it has been claimed that new sources of CP violation,
beyond that embedded in the CKM mass matrix,
will be required.

In fact, the most serious argument against spontaneous
CP violation probably comes from cosmology.
In a seminal paper, Zel'dovich, Kozbarev and
Okun \cite{domainwall} first pointed out that
spontaneously breaking of a discrete symmetry such as
CP in the early universe results in the
formation of domain walls during the phase
transition associated with the symmetry
breaking. Since these domain wall's total mass
is proportional to $\sigma R^2(t)$, where
$\sigma$ denotes the domain wall mass per unit
area and $R(t)$ denotes the cosmic scale
factor, their energy density
scales as $\approx 1/R(t)$.
In this case, the energy density of domain walls
can quickly come to dominate energy density in matter
and radiation, which scale as $1/R(t)^3$ and
$1/R(t)^4$ respectively.

One possible way out of a domain wall dominated
universe is to assume that the symmetry
breaking scale is set higher than the scale at
which inflation may occur so that domain
walls are diluted away during an inflatinary era.
In this case, feeding down CP violation to the
low-energy physics world requires some clever
model building \cite{nelsonbarr}.  On the other hand,
if we prefer the scale of
CP violation to be near the electroweak scale
for the purposes of baryogenesis, or to
explore possible new experimental signatures
at current or future accelerators,
this solution of the domain wall problem
is not available.

In this paper, we point out that because the
discrete CP symmetry is anomalous due to
the QCD axial anomaly, nonperturbative
communication between the fermion-Higgs
sector and the QCD sector leads to a tiny
but cosmologically significant splitting
of the CP conjugate vacuum degeneracy,
\sl if we assume a small but nonzero
$\theta_{QCD}$\rm (not $\bar \theta =
\theta_{QCD} + C_2 (R) Arg det M$)\footnote{Note
that most models of spontaneous CP breaking at
high energy scales \cite{nelsonbarr}
were previously designed so that $\bar \theta = 0$ at
tree level.}.
Therefore, as has been shown for such anomalous
discrete symmetries \cite{wiseetal}, one can avoid
the Zel'dovich \etal cosmological domain wall
problem. In the following, we will illustrate this
mechanism through a simplified model of spontaneous
CP breaking of two Higgs doublets.
However, the same mechanism should apply to more
realistic models.

Let us start with Lee's model \cite{tdlee} of two
Higgs doublets that conserves flavor. Neutral flavor
conservation (NFC) is achieved, for example, by imposing
Glashow-Weinberg's ${\bf Z}_2$ discrete symmetry
$$
\phi_1 \rightarrow \phi_1, \,\,\,\,\phi_2
\rightarrow - \phi_2, \,\,\,\,
U_r^o \rightarrow U_r^o, \,\,\,\,
D_r^o \rightarrow - D_r^o
\eqno (1)
$$
in which the quarks $U^o, D^o$ denote weak eigenstates.
The most general, renormalizable Higgs potential and
Yukawa interactions respecting the Glashow-Weinberg
${\bf Z}_2$ symmetry reads
$$
\eqalign{
L_{Higgs} =& -\mu_1^2 |\phi_1|^2 -\mu_2^2
|\phi_2|^2 \cr
           & + \lambda_1 |\phi_1|^4 + \lambda_2
|\phi_2|^4 + \lambda_3
|\phi_1|^2|\phi_2|^2 \cr
    & + \lambda_4 |\phi^\dagger_1 \phi_2|^2 +
\lambda_5 [(\phi_1^\dagger \phi_2)^2 +
(\phi_2^\dagger \phi_1)^2]}
\eqno (2)
$$
and
$$
L_{Yukawa} = (f_{ij} \bar Q_{Li}^o \tilde \phi_1 U_{Rj}^o
           + g_{ij} \bar Q^o_{Li} \phi_2 D_{Rj}^o  + h.c.).
\eqno (3)
$$
We assumed that the above terms are CP-invariant
and thus all the coupling
constants are purely real-valued.
As the electroweak symmetry is broken by the
vacuum expectation values,
$<\! \phi_1\!> \ne 0, \,\,\, <\!\phi_2\!> \ne 0$,
the Glashow-Weinberg discrete symmetry is also
spontaneously broken. Therefore cosmologically
dangerous ${\bf Z}_2$
domain walls arise at the phase transition.
However, this ${\bf Z}_2$ discrete symmetry
is anomalous due to
nonperturbative QCD effects \cite{wiseetal}.
In other words, the QCD instantons induce
an effective local operator
involving $2N_f$ quark flavors. This operator is
odd under the Glashow-Weinberg ${\bf Z}_2$ discrete symmetry,
which is thus explicitly broken. Preskill \etal     found that
the cosmological domain wall problem can disappear
due to this nonperturbative violation of the
${\bf Z}_2$ symmetry.
Note that these arguments remain valid irrespective
of whether the CKM matrix is chosen
to be real or not.

Alternatively, one may resort to an explicit but small
breaking of the Glashow-Weinberg ${\bf Z}_2$ discrete symmetry by
adding the following terms
$$
\delta L_{Yukawa} = \xi (\bar Q_{Li}^o f'_{ij}
\tilde \phi_2 U_{Rj}^o
+ \bar Q_{Li}^o g'_{ij} \phi_1 D_{Rj}^o + h.c.)
\eqno (4)
$$
and
$$
\delta L_{Higgs} = \xi' (\phi^\dagger_1 \phi_2 +
\phi_2^\dagger \phi_1)
(\alpha \phi^\dagger_1 \phi_1
+ \beta \phi^\dagger_2 \phi_2).
\eqno (5)
$$
This will solve the cosmological domain wall problem
associated with the Glashow-Weinberg's discrete symmetry,
without resorting to the QCD anomaly.
In addition, CP remains a manifest symmetry of
the Lagrangian. (This may also result in potentially
unacceptable flavor changing neutral-Higgs currents.
In the context of this toy model, however, we
will not concern ourselves about this problem.
Phenomenologically viable two Higgs doublet model \cite {lincoln2} or
models with a richer Higgs structure can presumably avoid it.)

However, a new domain wall problem apparently results
in this case. With the additional terms,
CP is spontaneously broken. This is because,
after the spontaneous electroweak symmetry breaking,
$$
\phi_1^o = {1 \over \sqrt 2} v_1 e^{i \theta_1},\hskip1cm
\phi_2^o = {1 \over \sqrt 2} v_2 e^{i \theta_2}
\eqno (6)
$$
in which the relative phase angle is
$$
\cos (\theta_1 - \theta_2)  \equiv cos (\kappa)
= -\xi' {\alpha v_1^2 +
\beta v_2^2 \over 4 \lambda_5 v_1 v_2}
\eqno (7)
$$
if we choose $\lambda_5 >0$.
As long as either one of $\xi$ and $\xi'$ is nonzero,
spontaneous weak-CP nonconservation arises.
Since weak CP is spontaneously broken,
a new kind of domain wall can result,
which separates two CP conjugate worlds in the
early Universe. Such domain walls are
cosmologically dangerous.

The cure is, interestingly enough,
connected with the QCD sector again,
{\it{provided that
a bare $\theta_{QCD}$ is nonzero and small}}
(this ugly feature is merely a restatement of the
`strong-CP' problem).
After the two Higgs fields get VEVs as in Eq.(6),
there is an induced flavor-diagonal CP violation,
$\theta_{QFD}$. The size of $\theta_{QFD}$ varies considerably
with the Yukawa coupling constants. If $\xi=0$, we find
$\theta_{QFD} (tree) \approx N_g (\theta_1 -
\theta_2)$ in which $N_g$ denotes the number of generations.
On the other hand,
if $\xi f' \approx g$ and $\xi g' \approx f$,
we find that $\theta_{QFD}
(tree) \approx 0$. Nevertheless, there is
generically a one-loop induced $\theta_{QFD}$ in
the latter case, and is conservatively estimated
to be $\theta_{QFD}(1 \,\, loop) \approx \xi'
{G_F \over 16 \pi^2} m_t^2 \approx
\xi' 10^{-4}$ for a top  quark mass
$m_t \approx 100GeV$.
With a reasonably small value of $\xi'$,
$\theta_{QFD} (1 \,\, loop)$ can be as small as $10^{-9}$.
Thus, in what follows, we assume that both $\theta_{QCD}$
and $\theta_{QFD}$ are typically of order $10^{-9}$ so that
the $\bar \theta \equiv \theta_{QCD} + \theta_{QFD}$
remains small $10^{-9}$.

At temperatures of the universe below the electroweak scale
but well above $\Lambda_{QCD}$, thermal suppression of QCD
instanton effects renders the domain wall practically stable.
The walls evolve and stretch out with expansion of the
universe. As the temperature approaches $\Lambda_{QCD}$,
instanton effects turn on and
yield a vacuum energy (in the zero temperature limit)
of order
$$
E_{vacuum} \approx \Lambda_{QCD}m_um_dm_s\cos \bar \theta.
\eqno (8)
$$
Under a $CP$ transformation, $\theta_{QFD} \rightarrow -
\theta_{QFD}$.
Therefore, the CP conjugate degenerate vacua are
now split in energy density by an amount
$$
\Delta E_{vacuum} \approx \Lambda_{QCD}m_um_dm_s
 \sin \theta_{QCD} \sin \theta_{QFD}.
\eqno (9)
$$
Therefore, domain walls created at the electroweak
phase transition begin to feel an energy difference between
the two sides of the wall. This yields a non-zero pressure
on the domain walls, and they begin to move as
the false vacuum decays to the true vacuum \cite{coleman}.

We can use the arguments of Preskill \etal \cite{wiseetal}
to estimate whether the above energy difference is enough
for the domain walls to decay away before they start to
dominate the energy density of universe. Reflections of
relativistic particles off of the domain walls can
result in an effective wall viscosity $\eta
\approx T^4$ at temperature $T$, producing a
dragging pressure $p \approx T^4 v$.
On the other hand, the curvature of the wall on a scale
$R(T)$ produces a pressure $\approx {\sigma \over R(T)}$,
which tends to straighten the wall.  Here $\sigma$ is the
wall tension, which is roughly given by the mass per
unit area of the wall. Thus, when
the curvature induced pressure equals the viscous drag,
irregularities on a given scale to be smoothed out,
as long as the time scale associated with motion on a
scale $R(T)$ is smaller than the Hubble time.
One finds the critical straightening length
scale $R_s(T) \approx {{\sqrt \sigma /G_N} \over T^3}$.
This is the minimal length on which wall segments will
straighten out.  Since the wall energy density
is $\rho_{wall} \approx \sigma/R_s(T)$, use of this
value for $R_s(T)$ results in the
largest value of wall energy which has to be
dissipated, and thus also
in the most conservative constraints on models.
One finds
$$ \eqalign{
{\rho_{wall} \over \rho_{rad}} & \approx
\sqrt {\sigma G_N} {1 \over T}\cr
                       & \approx 10^{-8}{1 \over
                         \sqrt \lambda} T_{EW}.}
\eqno (10)
$$
Thus, with a conservative value of the Higgs quartic
coupling constant $\lambda \approx 10^{-4}$,
the domain walls would start to dominate around
$T \approx 300 eV$.  At the nucleosynthesis scale,
for example, the walls provide
a negligible contribution to the total energy
density of the universe.

Let us see when the vacuum energy difference is large
enough to drive the walls from the true to the false
vacuum regions. The walls quickly move to the speed of
light once the pressure provided by the vacuum
energy difference is larger than the viscosity
$\Delta E_{vacuum} \ge T^4$.
Assuming $\theta_{QCD} \approx \theta_{QFD}
\approx 10^{-9}$ and $\Delta E_{vacuum} \approx
\Lambda_{QCD}m_um_dm_s \theta_{QCD} \theta_{QFD}$,
this happens when
$$
T_d \approx 10^{-5} \Lambda_{QCD} \approx 1 KeV.
\eqno (11)
$$
By the time the walls have reached the speed of light,
the regions of false vacuum are quickly driven away.
In fact, this could easily occur when the wall
velocity is much slower.  At the time they start to move,
the mean spacing between walls could be a
small fraction of the horizon size.
Even assuming relativistic velocities are required,
walls would be driven out before
they start to dominate the energy density of Universe,
as long as  $\bar \theta \ge 10^{-10-11}$.

Note that the cosmological scales involved are
{\it{quite}} interesting.  These domain walls could
remain in existence down to
temperatures (i.e of $O(KeV)$) where present day galaxy
sized regions first came inside the horizon.
They might thus provide seeds for galaxy formation which
might be relevant for large scale structure analyses.
One should also point out that while the domain walls might
contribute a small contribution to the total energy density
at the time they decay, their disappearance could result
in energetic particle production.
This could have interesting effects including possibly alter
light element abundances produced during big bang
nucleosynthesis (BBN) (i.e. see \cite{krausslindleysarkar}).
Because there are various possibilities, including
photo-dissociation of deuterium and helium, and also
energetic baryon production which might re-initiate some
BBN reactions, one must investigate in detail the
decay chain resulting from bubble wall collisions using
explicit models for spontaneous CP breaking in
order to make detailed predictions of what, if any effects
there might be.

The main unattractive feature in all of this is our
assumption that $\bar \theta$ is small, in the absence of any
dynamical mechanism to make this so.  Of course, a natural
solution to this strong CP problem is obtained by introducing
phenomenologically viable,
invisible axions.  One might hope that the introduction of a
Peccei-Quinn mechanism might
allow $\bar \theta$ to start out large (so that the domain
walls associated with spontaneous CP violation at the
electroweak scale might be quickly removed) and that at a
lower scale when axion mass effects turn on, $\bar \theta$
can relax to zero.  We find
that this cannot be easily achieved however.

As pointed out by Preskill et al \cite{wiseetal},
incorporating a Peccei-Quinn symmetry dynamically affects
the cosmology of domain walls as discussed above.
In the model introduced by Kim \cite{ksvz},
there is an extra coupling involving an
electroweak singlet heavy quark and a singlet Higgs field
$$ L_{kim} = g \bar Q_L Q_R
\Phi + c.c. \eqno (12)
$$
The weak isodoublet Higgs sector is largely unchanged, and
spontaneous CP violation can be accomodated as
discussed above.
However, since the Peccei-Quinn symmetry:
$Q_R \rightarrow e^{i \theta_{pq}} Q_R;
\,\, \Phi \rightarrow e^{-i \theta_{pq}} \Phi$
also suffers the same QCD anomaly as the discrete CP symmetry,
one can find a linear combination of these two
anomalous symmetries to yield a new discrete,
but nonanomalous symmetry.
(The other remaining anomalous continuous symmetry is the
Peccei-Quinn symmetry, which can still
solve the strong CP problem.) This symmetry then suffers
the standard domain wall problem when it is spontaneously
broken by the Higgs VEVs at the weak scale.

The other type of axion model due to Dine, Fischler and
Srednicki and Zhitniskii (DFSZ) \cite{dfsz}
is also problematic. They introduced a singlet scalar
field $\Sigma$:
$$
L_{DFSZ} = \lambda_{PQ} \phi_1 \phi_2^\dagger
\Sigma^2 + h.c.
\eqno (13)
$$
The Peccei-Quinn symmetry breaking
$<\! \Sigma \!> = v_{PQ}$ gives
rise to a term in the low-energy effective Lagrangian
$ \lambda_{PQ} v_{PQ}^2 \phi_1 \phi_2^\dagger + h.c.$.
However, the Higgs potential and Yukawa coupling in
Eqs. (2-5) do not
respect the Peccei-Quinn symmetry as long as any of
$\lambda_5, \xi, \xi'$ remain nonzero.
One would be required to set $\xi=0$, and somehow
fine tune $\lambda_5, \,\, \xi' \rightarrow 0$,
while keeping their ratio fixed, in
order for the terms leading to a possible spontaneous
CP violation at the weak scale (see eq. (8)) not to
also violently break the Peccei-Quinn
symmetry.  Thus, barring an apparently unnatural fine tuning,
in the DFSZ axion model, it seems that
the only viable option of CP violation is through the
CKM mass matrix.

Are there extensions of the DFSZ axion models that
accomodate spontaneous CP violation
at the weak scale? Extensions involving either two
isodoublets and two isosinglets
or three isodoublets and one isosinglet
one might still lead to non-trivial CP violating Higgs VEV
phases without introducing extra non-anomalous discrete
symmetries, or simultanously explicitly breaking the PQ
symmetry. Extensions of Higgs sector beyond the two Higgs
models
are also necessary to be phenomenologically realistic (e.g.
recall the problem of flavor changing neutral currents).
Such extended models may have several new interesting
features and are currently under study.

SJR acknowledges
the hospitality of the Institute for Theoretical Physics at
Santa Barbara during the course of this work.
This work was supported in part by the National Science
Foundation under Grant No. PHY89-04035.

\bibliographystyle {unsrt}

\end{document}